\documentclass[aps,prd,superscriptaddress,twoside,twocolumn,nofootinbib,10pt,%
showpacs,floatfix]{revtex4-1}

\usepackage{amsmath,amssymb}
\usepackage{graphicx,bm}
\usepackage{slashed}
\usepackage{dcolumn}
\usepackage{epstopdf}
\usepackage{ulem} 
\usepackage[usenames]{color}

\allowdisplaybreaks

\usepackage{graphicx}
\usepackage{amssymb}
\usepackage{epstopdf}
\usepackage{bm}
\usepackage{ulem} 
\usepackage[usenames]{color}

\renewcommand\sout{\bgroup \color{red} \ULdepth=-.5ex \ULset}

\begin{document}

\title{Chiral partner structure of heavy baryons
from the bound state approach with hidden local symmetry}

\author{Masayasu Harada}

\author{Yong-Liang Ma}

\affiliation{Department of Physics, Nagoya University, Nagoya, 464-8602, Japan.}
\date{\today}
\begin{abstract}
The chiral partner structure of heavy baryons is studied using the
bound state approach by binding the heavy-light mesons to the nucleon as
the soliton in an effective Lagrangian for the pseudoscalar and vector
mesons based on hidden local symmetry. In the heavy-light meson sector, we regard the $H$ doublet and $G$ doublet as chiral partners and couple them to light mesons with a minimal derivative. We find that the chiral
partner of $\Lambda_c(\frac{1}{2}^+,2286)$ 
is the $\Lambda_c(\frac{1}{2}^-,\frac{3}{2}^-)$ heavy quark doublet with a mass of about
$3.1$\,GeV but not ($\Lambda_c(\frac{1}{2}^-,2595)$,
$\Lambda_c(\frac{3}{2}^-,2625)$), which might be
interpreted as an orbital excitation of
$\Lambda_c(\frac{1}{2}^+,2286)$. The same model is applied to the
bottom sector, and the chiral partner of
$\Lambda_b(\frac{1}{2}^+,5625)$ is shown to have a mass of about
$6.5$\,GeV. We also discuss the chiral partner structures for the isospin vector
heavy baryons. For the pentaquark states, we find that 
the masses of the pentaquark state made of a ground state heavy-light meson and its chiral partner are similar, and both of them are below the $Dp$ threshold, which therefore cannot be ruled out by the present data.
\end{abstract}

\maketitle

\section{Introduction}

Chiral symmetry plays an important role in hadron physics. When we
set $N_f$ flavor light quarks to be massless, quantum chromodynamics (QCD) has an $SU(N_f)_L
\times SU(N_f)_R$ chiral symmetry at the Lagrangian level. The chiral
symmetry is not preserved by QCD vacuum but broken dynamically to
its vector component, i.e., $SU(N_f)_L \times SU(N_f)_R \rightarrow
SU(N_f)_V$. This dynamical chiral symmetry breaking splits the
degeneracy of the chiral parters, which are supposed to be
degenerate when the chiral symmetry is restored so that the study
of the chiral partner structure of hadrons can help us reveal the
magnitude of the chiral symmetry breaking, i.e., the order
parameter.

In the light meson sector, the chiral partner structure might be
complicated since two light quarks are involved (see, e.g., Refs.~\cite{Weinberg:1990xn,Harada:2003jx}). On the other hand, studying the chiral partner structure in the heavy-light meson sector would be
easier since they include only one light quark, the dynamics of
which is controlled by the chiral symmetry. In addition to the
chiral symmetry, the dynamics of the heavy-light mesons is also
controlled by spin-flavor symmetry due to their heavy quark constituent~\cite{Wise:1993wa}. Based on this heavy quark symmetry, the ground states form a doublet with spin-parity quantum numbers $(1^-,
0^-)$, and the first excited states belong to the $(1^+,0^+)$ doublet.
In Ref.~\cite{heavy-partner}, it was proposed that these two
doublets were chiral partners to each other in the QCD-like models. As
a signature of the chiral partner structure, the mass splitting of
them is induced by the dynamical breaking of the chiral symmetry so
that the mass difference is about the constituent quark mass. This
was confirmed by the spectrum of the relevant particles;
$m_{D_0^{\ast}} - m_D \simeq m_{D_1}- m_{D^{\ast}} \simeq 450~$MeV
is at the same order of $m_{D_{s0}}(2317) - m_{D_s} \simeq
m_{D_{s1}}(2460)- m_{D_{s}^{\ast}} \simeq 350~$MeV (see, e.g., Refs.~\cite{heavy-mass-diffe1,heavy-mass-diffe2}).

In Ref.~\cite{Nowak:2004jg}, the analysis on the chiral partner
structure of heavy baryons was made based on the bound state
picture~\cite{Callan:1985hy,bound-state-approach} together with the
heavy quark symmetry in which the heavy baryon is
introduced~\cite{heavy-baryon-bound-state1,Gupta-Momen-Schechter-Subbaraman,Oh:1994ky,Nowak:2004jg,Harada:1997bk,Wu:2004wg,heavy-baryon-bound-state2}
as the heavy mesons bound with the nucleon as the soliton. In the
analysis, the excited heavy baryon $\Lambda_c(2595)$ with $J^P =
\frac{1}{2}^-$ is regarded as the chiral partner to the ground state
baryon $\Lambda_c(2268)$ ($J^P = \frac{1}{2}^+$).

In this paper, we revisit the chiral partner structure of the heavy
baryons in the bound state approach based on our recent progress in the soliton property~\cite{MOYH, MOYHLPR} and the effective Lagrangian for the heavy-light meson chiral partner structure~\cite{HHM1,HHM2}. In the light meson sector, we considered all the $\mathcal{O}(N_c)$ terms of hidden local symmetry (HLS), all the $\mathcal{O}(p^2)$, $\mathcal{O}(p^4)$, and homogeneous Wess-Zumino (hWZ) terms~\cite{Harada:2003jx,Bando:1984ej}. In the heavy and light meson interaction sector, we start with the interaction Lagrangian for the heavy-light meson and
light mesons analyzed in Refs.~\cite{HHM1,HHM2}, 
where the chiral partner is introduced in
the framework of a linear sigma model. We integrate out the scalar mesons and integrate in the vector mesons
to construct an effective Lagrangian for heavy mesons interacting with
the pseudoscalar mesons and vector mesons based on the HLS~\cite{Bando:1984ej,Harada:2003jx} (see, e.g., Refs.~\cite{HLS-alter} for alternative approaches), and
the heavy quark symmetry. Then, we consider that the static soliton
couples to the heavy-light meson and study their spectrum. After the
derivation of the heavy baryon spectrum in the static case, we
consider the collective coordinate quantization to make states
definite quantum numbers. Our explicit calculation shows that, in
the heavy quark limit and large $N_c$ limit, up to the $\mathcal{O}(p^4)$ terms of HLS, the chiral partner of
$\Lambda_c(\frac{1}{2}^+,2286)$ is predicted to be the $\Lambda_c(\frac{1}{2}^-,\frac{3}{2}^-)$ heavy quark doublet 
with a mass of about $3.1$\,GeV but not ($\Lambda_c(\frac{1}{2}^-,2595)$,
$\Lambda_c(\frac{3}{2}^-,2625)$) listed in the PDG
table~\cite{Nakamura:2010zzi}, which might be interpreted as an
orbital excitation of $\Lambda_c(\frac{1}{2}^+,2286)$. 
Extending our approach to the bottom sector we predicted the chiral partner structure of bottom baryons. We finally studied the pentaquark spectrum using our framework and found that the masses of the pentaquark states made of a ground state heavy-light meson and its chiral partner are similar and both of them are below the $Dp$ threshold, which therefore cannot be ruled out by
the present data~\cite{Pentadata}.

This paper is organized as follows: 
In Sec.~\ref{sec:effecLHeavy}, the chiral partner structure of heavy
baryons is studied. We derive the analytic forms of the heavy
baryons' masses. The heavy baryon spectrum with chiral partner
structure is estimated in Sec.~\ref{sec:heavybaryon}, and the pentaquark state's spectrum is estimated in Sec.~\ref{sec:penta}. The last section is for a summary and discussions. Some useful explicit derivations are given in the Appendix.

\section{Heavy Baryons in the Effective Lagrangian
for Heavy-light Mesons with Chiral Doubling}

\label{sec:effecLHeavy}

\subsection{Effective Lagrangian for heavy-light mesons
with chiral doubling}

Here, we construct the effective Lagrangian describing the
interaction between the heavy-light mesons and the light mesons.
With respect to the chiral transformation property of the light
quarks in the heavy-light mesons, the heavy-light meson field can be
decomposed into a right-handed component $\mathcal{H}_R$ and
left-handed one $\mathcal{H}_L$~\cite{heavy-partner}. Under
chiral $SU(2)_L \times SU(2)_R$ symmetry, they transform as
\begin{eqnarray}
\mathcal{H}_L & \to & \mathcal{H}_L g_L^\dag, \;\;\;\;\;
\mathcal{H}_R \to \mathcal{H}_R g_R^\dag.
\end{eqnarray}
In Refs.~\cite{HHM1,HHM2} these fields are used to construct a
Lagrangian where the chiral symmetry are realized linearly. 
In the present paper, for the study of the heavy baryon spectrum 
in the bound state approach, we adopt the 
nonlinear realization of the chiral symmetry.
Then, by replacing $M$ in Eq.~(29) of Refs.~\cite{HHM1,HHM2} 
with $F_\pi U$ where $U = e^{2 i \pi/F_\pi}$, we obtain
\begin{eqnarray}
\mathcal{L}_{\rm heavy} & = & \frac{1}{2} {\rm Tr}\left[
\bar{\mathcal{H}}_L i(v \cdot \partial) \mathcal{H}_L\right] 
+
\frac{1}{2}{\rm Tr} \left[\bar{\mathcal{H}}_R i(v\cdot
\partial)\mathcal{H}_R\right] \nonumber\\
& & {} 
-\frac{\Delta}{2} {\rm Tr}\left[\bar{\mathcal{H}}_L \mathcal{H}_L
+ \bar{\mathcal{H}_R} \mathcal{H}_R\right] 
\nonumber\\
& & {} -\frac{g_\pi F_\pi}{4}{\rm Tr}\left[ U^\dagger
\bar{\mathcal{H}}_L\mathcal{H}_R 
+ U
\bar{\mathcal{H}}_R\mathcal{H}_L\right] 
\nonumber\\
&& {}
+ i\frac{g_A}{2} 
{\rm Tr}\left[ \gamma^5 \gamma^\mu \partial_\mu  U^\dagger
  \bar{\mathcal{H}}_L \mathcal{H}_R \right. \nonumber\\
  & & \left. {} \qquad \qquad \; -
  \gamma^5 \gamma^\mu \partial_\mu U \bar{\mathcal{H}}_R
  \mathcal{H}_L \right] , 
\label{eq:heavylagranchiral}
\end{eqnarray}
where $\Delta, g_\pi$ and $g_A$ are parameters.
In the present work, we include the vector mesons using the 
HLS~\cite{Bando:1984ej,Harada:2003jx} by introducing
the matrix valued variables $\xi_L$ and $\xi_R$ 
as $U = \xi_L^\dag \xi_R$. Then, similarly to Ref.~\cite{heavy-mass-diffe2} we
convert the heavy meson fields as
\begin{eqnarray}
\hat{\mathcal{H}}_L & = & \mathcal{H}_L \xi_L^\dag, \;\;\;\;\;
\hat{\mathcal{H}}_R = \mathcal{H}_R \xi_R^\dag,
\label{eq:chiralphys}
\end{eqnarray}
which under the full symmetry transformation $G_{\rm full} =
[SU(2)_L\times SU(2)_R]_{\rm chiral} \times [U(2)]_{\rm HLS}$
transform as
\begin{eqnarray}
\hat{\mathcal{H}}_L \rightarrow \hat{\mathcal{H}}_L h^\dag(x)\ ,
\quad
\hat{\mathcal{H}}_R \rightarrow \hat{\mathcal{H}}_R h^\dag(x)\ .
\end{eqnarray}
Associated with the field redefinitions in Eq.~(\ref{eq:chiralphys}), 
it is convenient to use the following quantities for the $\pi$ fields:
\begin{eqnarray}
\hat{\alpha}_{\parallel\mu} & = & \frac{1}{2i}\left(D_\mu \xi_R
\cdot
\xi_R^\dag + D_\mu \xi_L \cdot \xi_L^\dag \right), \nonumber\\
\hat{\alpha}_{\perp\mu} & = & \frac{1}{2i}\left(D_\mu \xi_R \cdot
\xi_R^\dag - D_\mu \xi_L \cdot \xi_L^\dag \right)\ ,
\label{eq:1-form}
\end{eqnarray}
where the covariant derivative $D_\mu$ is given by
$D_\mu = \partial_\mu - i V_\mu$ with $V_\mu$ being the gauge 
field of the HLS. By using these quantities, the above Lagrangian 
is extended to include the vector mesons as
\begin{eqnarray}
{\cal L}_{\rm heavy} & = & 
\frac{1}{2} {\rm Tr }
\Big[\hat{\mathcal{H}}_L(i v\cdot \widetilde{D}) \bar{\hat{\mathcal{H}}}_L \Big] 
+ \frac{1}{2} {\rm Tr } \Big[\hat{\mathcal{H}}_R(i v\cdot \widetilde{D}) \bar{\hat{\mathcal{H}}}_R \Big] \nonumber\\
& & {} -\frac{\Delta}{2} {\rm Tr}\left[ \bar{\hat{\mathcal{H}}}_L \hat{\mathcal{H}_L } + \bar{\hat{\mathcal{H}_R}} \hat{\mathcal{H}_R} \right] 
\nonumber\\
&& {} -\frac{g_\pi F_\pi}{4}{\rm Tr}\left[ 
\bar{\hat{{\mathcal{H}}}}_L \hat{\mathcal{H}}_R
+ \bar{\hat{\mathcal{H}}}_R\hat{\mathcal{H}}_L\right] \nonumber\\
& & {} - g_A 
{\rm Tr}\left[ \gamma^5 \gamma^\mu \hat{\alpha}_{\perp\mu} \left( \bar{\hat{\mathcal{H}}}_L \hat{\mathcal{H}}_R + \bar{\hat{\mathcal{H}}}_R \hat{\mathcal{H}}_L \right)
\right] , 
\label{eq:effectheavychiral}
\end{eqnarray}
where $\widetilde{D}$ is defined as
$\widetilde{D}_\mu = \partial_\mu - i V_\mu - i \kappa \alpha_{\parallel\mu}$ with $\kappa$ being a
real parameter measuring the magnitude of the violation of the vector meson dominance.

To study the chiral partner structure of the heavy baryons, we
rewrite the Lagrangian (\ref{eq:effectheavychiral}) in terms of the
heavy-light meson doublets $\hat{H}$ and $\hat{G}$ with quantum
numbers $\hat{H} = (0^-,1^-) $ and $\hat{G} = (0^+,1^+)$;
specifically, we make the substitution
\begin{eqnarray}
\hat{\mathcal{H}}_L & = & \frac{1}{\sqrt{2}}[\hat{G} - i \hat{H}
\gamma_5], \;\;\;\;\; \hat{\mathcal{H}}_R =
\frac{1}{\sqrt{2}}[\hat{G} + i \hat{H} \gamma_5].\label{eq:heavysub}
\end{eqnarray}
In terms of the physical states, the $\hat{H}$ and $\hat{G}$
doublets can be explicitly expressed as
\begin{eqnarray}
\hat{H} & = & \frac{(1+v\hspace{-0.2cm}\slash)}{2}[D^{\ast \, ;
\mu}\gamma_\mu + i D \gamma_5], \nonumber\\
\hat{G} & = &
\frac{(1+v\hspace{-0.2cm}\slash)}{2}[ - D^{\prime \, ;
\mu}_{1}\gamma_\mu\gamma_5 + D_0^{\ast} \,].\label{eq:hgphys}
\end{eqnarray}

Substituting Eq.~(\ref{eq:heavysub}) into Eq.~(\ref{eq:effectheavychiral}), we obtain
\begin{eqnarray}
{\cal L}_{\rm heavy} & = & 
- {\rm Tr}\Big[ \hat{G} ( i v\cdot \widetilde{D} ) \bar{\hat{G}} \Big] 
+ {\rm Tr}\Big[ \hat{H} (i v\cdot \widetilde{D})\bar{\hat{H}} \Big] \nonumber\\
& & {} - \frac{\Delta}{2}\,
  {\rm Tr}\Big[\hat{G}\bar{\hat{G}} - \hat{H}\bar{\hat{H}}\Big] 
- \frac{g_\pi F_\pi}{4} 
  {\rm Tr}\Big[ \hat{H}\bar{\hat{H}} + \hat{G}\bar{\hat{G}} \Big] 
\nonumber\\
& & {} + g_A {\rm Tr }\Big[
  \hat{H}\gamma_\mu\gamma_5 \hat{a}_{\perp}^\mu \bar{\hat{H}} \Big]
- g_A {\rm Tr }\Big[ \hat{G}\gamma_\mu\gamma_5
\hat{a}_{\perp}^\mu \bar{\hat{G}} \Big] 
\ .
\label{eq:effectheavyphys}
\end{eqnarray}
This expression explicitly shows that the $g_\pi F_\pi$ term splits the
spectrum of $\hat{H}$ and $\hat{G}$ doublets while the $\Delta$ term
shifts the masses of these two doublets toward the same direction.
In the present paper, we use the physical masses of heavy mesons as inputs to 
calculate the heavy baryon masses so that we drop the $g_\pi F_\pi$
term and the $\Delta$ term in the following calculation of the masses of heavy baryons. Note that due to the chiral partner structure adopted here, the magnitudes of the coupling constants in the last two terms of Eq.~(\ref{eq:effectheavyphys}) are the same; therefore, the chiral partner spectrum is predictable.

\subsection{Heavy baryon masses from the bound state approach}

In this subsection we derive the heavy baryon masses based on the
bound state approach~\cite{Oh:1994ky,Gupta-Momen-Schechter-Subbaraman,Nowak:2004jg,
Harada:1997bk,Wu:2004wg,heavy-baryon-bound-state2}.

To make the mesonic theory a baryonic one, we follow the
standard procedure to take the Hedgehog ansatze for a
classical soliton~\cite{Skyrme}
\begin{eqnarray}
\xi_{R} = \xi_{ L}^\dag =
\xi_c(\mathbf{x}) = 
\exp\left[i\bm{\tau}\cdot\hat{\mathbf{x}}\frac{F(x)}{2}\right]
\ ,
\label{eq:hedgehog}
\end{eqnarray}
with $\tau_i$ as the Pauli matrices and the subscript $c$ standing
for the classical solution. From the Hedgehog ansatze
(\ref{eq:hedgehog}), one can easily see that $\xi_c$ 
transforms under separate spatial rotation and isospin rotation but is
invariant under the combined rotation; i.e., the hedgehog profile
correlates the angular momentum and the isospin. For the vector
mesons, their profile functions can be parametrized
as~\cite{Jain:1987sz,para-vevtor}
\begin{eqnarray}
\omega_{\mu,c} & = & \omega(x)\delta_{0\mu}, \;\; \rho_{i,c}^a =
\frac{1}{gx}\epsilon_{ija}\hat{\mathbf{x}}_j G(x), \;\;
\rho^a_{0,c} = 0 .
\label{eq:ansatzv}
\end{eqnarray}
From the hedgehog ansatze (\ref{eq:hedgehog}) and profile
functions (\ref{eq:ansatzv}) we express the quantities
$\hat{\alpha}_\perp^{\mu}$ and $\hat{\alpha}_\parallel^{\mu}$ as
\begin{eqnarray}
\hat{\alpha}_{\perp}^\mu  = \left(0,\bm{a}_\perp \right) \ , \quad
\hat{\alpha}_{\parallel}^\mu = \left(
a_\parallel\,,\,\bm{a}_\parallel\right) \, ,
\label{eq:hedg1forms}
\end{eqnarray}
where
\begin{eqnarray}
\bm{a}_\perp &=&
\frac{1}{2}\left[
  \frac{\sin F(x)}{x} \bm{\tau} + 
  \left( F^\prime(x) - \frac{\sin F(x)}{x} \right)
  \left(\bm{\tau}\cdot \hat{\mathbf{x}} \right)
\hat{\mathbf{x}}\right]
\nonumber\\
a_\parallel &=& -\frac{g}{2}\omega(x)\ ,\nonumber\\
\bm{a}_\parallel &=& 
  \left[ \frac{1}{x}\sin^2\frac{F}{2} - \frac{1}{2x}G(x) \right] 
  \hat{\mathbf{x}} \times \bm{\tau}
\ .
\label{eq:coeffaparaperpen}
\end{eqnarray}

In the rest frame of the heavy-light meson, i.e., $v_\mu = (1, \bm{0})$,
the $\hat{H}$ doublet has nonvanishing elements only in the
upper-right $2 \times 2$ sub-block while the $\hat{G}$ doublet has
nonvanishing elements only in the upper-left $2 \times 2$ sub-block.
The matrix forms of $\hat{H}$ and $\hat{G}$ doublets become
\begin{eqnarray}
\hat{H} & = & \left(
          \begin{array}{cc}
            0 & \mathbb{H} \\
            0 & 0 \\
          \end{array}
        \right)
\ , 
\hat{G} = \left(
          \begin{array}{cc}
            \mathbb{G} & 0 \\
            0 & 0 \\
          \end{array}
        \right)
\ , \nonumber\\
\bar{\hat{H}} & = & \left(
          \begin{array}{cc}
            0 & 0 \\
            -\mathbb{H}^\dag & 0 \\
          \end{array}
        \right)
\ , 
\bar{\hat{G}} = \left(
          \begin{array}{cc}
            \mathbb{G}^\dag & 0 \\
            0 & 0 \\
          \end{array}
        \right)
\ . 
\label{eq:eomhg}
\end{eqnarray}
Then, the Lagrangian (\ref{eq:effectheavyphys}) is reduced to
\begin{eqnarray}
{\cal L}_{\rm heavy} & = & 
{} - {\rm Tr }\Big[ \mathbb{G} i \partial_0 \mathbb{G}^\dag \Big] 
- {\rm Tr }\Big[ \mathbb{H} i \partial_0 \mathbb{H}^\dag \Big] 
\nonumber\\
& & {} - \frac{1}{2} (1 + \kappa)\, g \omega(r)
  {\rm Tr }\Big[\mathbb{G}\mathbb{G}^\dag\Big] \nonumber\\
& & {} - \frac{1}{2} (1 + \kappa) \, g\omega(r) 
  {\rm Tr }\Big[ \mathbb{H}\mathbb{H}^\dag \Big] 
\nonumber\\
&&  {}
- g_A {\rm Tr }\Big[ \mathbb{H}\bm{\sigma} \cdot
  \bm{a}_\perp \mathbb{H}^\dag \Big] 
+ g_A {\rm Tr }\Big[ \mathbb{G}\bm{\sigma} \cdot \bm{a}_\perp
  \mathbb{G}^\dag \Big] .
\label{eq:effectheavyphys2c}
\end{eqnarray}

Since the hedgehog ansatz for the Skyrme soliton correlates
the angular momentum and isospin, the bound states should be
invariant under the ``grand spin'' rotation with the operator
defined as
\begin{eqnarray}
\mathbf{G} & = & \mathbf{r} + \mathbf{J} + \mathbf{I}_{\rm light},
\end{eqnarray}
where $\mathbf{r}, \mathbf{J}$, and $\mathbf{I}_{\rm light}$ are the
ordinary orbital angular momentum between the soliton and
heavy-light meson, heavy meson spin, and the heavy meson isospin
operators. Taking into account that the heavy quark spin is
conserved in the heavy quark limit, one simply defines the ``light
quark grand spin" operator
\begin{eqnarray}
\mathbf{g} & = & \mathbf{r} + \mathbf{J}_{\rm light} +
\mathbf{I}_{\rm light},
\end{eqnarray}
with $\mathbf{J}_{\rm light}$ as the spin operator of the light
degree of freedom of the heavy-light meson, and in both $H$ and $G$
doublets, the eigenvalue of the operator $\mathbf{J}_{\rm light}$ is
$1/2$ so that the eigenmodes of the heavy baryons can be classified
by the third component of heavy quark spin $s_Q$ and the light quark
grand spin $(g,g_3)$ and the parity ${\rm P}$.

Taking into account the isospin, light quark spin and heavy quark
spin indices that the heavy-light meson has, one can write the
static wave functions of the heavy-light mesons
as~\cite{Gupta-Momen-Schechter-Subbaraman}
\begin{eqnarray}
\mathbb{H}^{\dag,a}_{c,lh} & = & u^{(H)}(\mathbf{x})({\bm \tau} \cdot
\hat{\mathbf{x}})_{ad}
\psi_{dl}^{(H)}(g,g_3;r,k) \chi_h^{(H)}
\ ,
\nonumber\\
\mathbb{G}^{\dag,a}_{c,lh} & = & u^{(G)}(\mathbf{x})({\bm \tau} \cdot
\hat{\mathbf{x}})_{ad}
\psi_{dl}^{(G)}(g,g_3;r,k) \chi_h^{(G)} ,
\label{eq:eigenggen}
\end{eqnarray}
where $a, l$, and $h$ denote the indices for the isospin of the
heavy-light meson, the spin of the light degree of freedom, and the
heavy quark spin, respectively. 
$k$ is the eigenvalue of the operator 
\begin{equation}
\mathbf{K} = \mathbf{I}_{\rm light} + \mathbf{J}_{\rm light} ,
\end{equation}
with $k_3$ as its third component. $\chi_h^{(H,G)}$ is factorized out due
to the conservation of the heavy quark spin. $u(x)$ is a radial
function which is strongly peaked at the origin and normalized as
$\mathbf{x}^2 |u(\mathbf{x})|^2 \simeq \delta^3(\mathbf{x})
$~\cite{Gupta-Momen-Schechter-Subbaraman,Oh:1994ky}. 
This implies that the relevant matrix elements are independent of the
quantum number $r$~\cite{Harada:1997bk}. The generalized ``angular"
wave function $\psi_{dl}^{(H,G)}(g,g_3;r,k)$ can be expanded
as~\cite{Harada:1997bk}
\begin{eqnarray}
\psi_{dl}^{(H,G)}(g,g_3;r,k) & = &
\sum_{r_3,k_3}C_{r_3,k_3;g_3}^{~r,k;g}Y_{r}^{r_3}\xi_{dl}(k,k_3),
\label{eq:heavywave}
\end{eqnarray}
where $Y_r^{r_3}$ stands for the standard spherical harmonic representing
orbital angular momentum $r$ while $C$ denotes the
ordinary Clebsch-Gordan coefficients. $\xi_{dl}(k,k_3)$
represents a wave function in which the ``light spin'' and ``light isospin'' referring
to the ``light cloud'' component of the heavy meson are
added vectorially to give $\mathbf{K} = \mathbf{I}_{\rm light} + \mathbf{J}_{\rm light}$
with eigenvalues $\mathbf{K}^2 = k(k+1)$.
Note that in the present analysis, $k$ is a good quantum number since the
relevant matrix elements are independent of the quantum number $r$.
Furthermore, both the quantum numbers for 
the ``light spin'' and ``light isospin'' are given by $1/2$ so that
the possible values of $k$ are either $0$ or $1$.
The normalization of the eigenstate $\psi_{dl}^{(H,G)}(g,g_3;r,k)$ is
\begin{eqnarray}
& & \int d\Omega \psi_{dl}^{(H,G)}(g,g_3;r,k)
\left[ \psi_{d^\prime l^\prime}^{(H,G)} 
 \left(g^\prime,g_3^\prime;r^\prime,k^{\prime}\right) \right]^\dag \nonumber\\
& & \quad = 
\delta_{dd^\prime}\delta_{ll^\prime}\delta_{gg^\prime}\delta_{g_3g_3^\prime}\delta_{rr^\prime}\delta_{k
k^{\prime}} , \label{eq:normkspin}
\end{eqnarray}
where $\int d\Omega$ is the solid angle integration.

From Lagrangian (\ref{eq:effectheavyphys2c}), we parametrize
the potential as
\begin{eqnarray}
V & = &\left(
         \begin{array}{cc}
           V_{\rm H} & 0 \\
           0 & V_{\rm G} \\
         \end{array}
       \right)\, .
\end{eqnarray}
By substituting the ansatz (\ref{eq:heavywave}),
$V_H$ and $V_G$ are obtained as
\begin{eqnarray}
V_{\rm H} 
&=&
\frac{1}{2} (1 + \kappa ) \, g \omega(r) 
  {\rm Tr }\Big[ \mathbb{H}_c\mathbb{H}_c^\dag \Big]
+ g_A {\rm Tr }\Big[ 
    \mathbb{H}_c\bm{\sigma} \cdot \bm{a}_\perp \mathbb{H}_c^\dag \Big]
\nonumber\\
& = & \frac{1}{2} (1+\kappa)\,g \omega(0) 
+ g_A  F^\prime(0) \Big[k(k+1) - \frac{3}{2}\Big]
\ ,
\nonumber\\
V_{\rm G}
& = & 
\frac{1}{2} (1 + \kappa) \, g \omega(r)
  {\rm Tr }\Big[ \mathbb{G}_c\mathbb{G}_c^\dag\Big] 
- g_A {\rm Tr }\Big[ \mathbb{G}_c\bm{\sigma} \cdot 
  \bm{a}_\perp \mathbb{G}_c^\dag \Big] 
\nonumber\\
& = & \frac{1}{2} (1+\kappa)\,g \omega(0) 
- g_A F^\prime(0)\Big[k(k+1) - \frac{3}{2}\Big]
\ .
\label{eq:bindenergy}
\end{eqnarray}

Next, we make a quantization by
a time dependent $SU(2)$ rotation of the fields in the HLS
Lagrangian in the light-quark sector as
\begin{eqnarray}
\xi_c(\mathbf{x}) & \rightarrow & \xi(\mathbf{x},t) = C(t)
\xi_c(\mathbf{x}) C^\dag(t), \nonumber\\
V_{\mu,c}(\mathbf{x})
& \rightarrow & V_{\mu}(\mathbf{x},t) = C(t) V_{\mu,c}(\mathbf{x})
C^\dag(t),\label{eq:mesoncollective}
\end{eqnarray}
where $C(t)$ is a time dependent unitary matrix satisfying
$C(t)C(t)^\dag = C(t)^\dag C(t) = 1$. 
Accordingly, the heavy-light meson fields are rotated as
\begin{eqnarray}
\mathbb{H}(\mathbf{x},t) = \mathbb{H}_{c}(\mathbf{x}) C^\dag(t)
\ ,\quad
\mathbb{G}(\mathbf{x},t) = \mathbb{G}_{c}(\mathbf{x}) C^\dag(t)
\ .
\label{eq:heavycollective}
\end{eqnarray}

This collective rotation gives an additional contribution to the
Lagrangian
\begin{eqnarray}
\delta{\cal L}_{\rm coll} & = & \frac{1}{2} \mathcal{I}
\Omega^2 + \mathbf{I}_{\rm light} \cdot
\bm{\Omega}
\ ,
\label{eq:hisospin}
\end{eqnarray}
where the angular velocity $\Omega_i$ corresponding to the collective coordinate rotation
is defined as
\begin{eqnarray}
\frac{1}{2}i \bm{\tau} \cdot \bm{\Omega} & \equiv & C^\dag
\partial_0 C
\ .\label{eq:angularvelocity}
\end{eqnarray}
$\mathcal{I}$ is the moment of the inertia of the
soliton configuration.
By using $\mathcal{I}$, the light baryon masses are expressed as
\begin{eqnarray}
m_b & = & M_{\rm sol} + \frac{j_b(j_b+1)}{2\mathcal{I}}
\ ,
\end{eqnarray}
where $M_{\rm sol}$ is the soliton mass and $j_b$ is the spin of light baryon.
Using the nucleon mass $m_N$ and the delta mass $m_\Delta$ as inputs, $M_{\rm sol}$ and $\mathcal{I}$ are given as
\begin{eqnarray}
M_{\rm sol} & = & \frac{5 m_N - m_\Delta}{4} \ , \qquad \mathcal{I} = \frac{3}{2\left( m_\Delta - m_N \right)}
\ .\label{eq:inemohls}
\end{eqnarray}

From Eq.~(\ref{eq:hisospin}) one obtains~\cite{Callan:1985hy}
\begin{eqnarray}
\delta{\cal L}_{\rm coll} & = & \frac{1}{2} \mathcal{I}
\Omega^2 - \chi(k)\mathbf{K} \cdot \bm{\Omega} , \label{eq:Lrotate}
\end{eqnarray}
where the coefficient $\chi(k)$ is calculated 
as
\begin{eqnarray}
\chi(k) = \left\{
\begin{array}{ll}
0 \ , & \ (k=0) \ , \\
\frac{[k(k + 1) + 3/4 - j_l(j_l+1)]}{2k(k+1)}\ , 
 & (k\neq0) \\
\end{array}
\right.
\ ,
\end{eqnarray}
with $j_l$ being the spin quantum number of the light degree of 
freedom of tne heavy-light meson.
For convenience we show the derivation in the Appendix.
The Hamiltonian of the collective rotated system is obtained by
the standard Legendre transformation as
\begin{eqnarray}
H_{\rm coll} & = & \frac{1}{2\mathcal{I}}
[\mathbf{J}^{\rm sol} + \chi(k)\mathbf{K}]^2, \label{eq:hrotate}
\end{eqnarray}
where $\mathbf{J}^{\rm sol}$ is the canonical momentum conjugating
to the collective variable $C(t)$:
\begin{eqnarray}
{J}_a^{\rm sol} & = & \frac{\partial \delta {\cal L}^{\rm coll}}
{\partial\Omega^a} = \mathcal{I} \Omega_a + I_{\rm light}^a.
\end{eqnarray}
The first term $\mathcal{I} \Omega_a$ is the isospin
operator for the light baryon sector while the second term $I_{\rm
light}^a$ is the isospin operator for the heavy-light mesons
interacting with the light baryon so that $\mathbf{J}^{\rm sol}$ is
identical to the isospin operator for the heavy baryon $\mathbf{I}$:
\begin{equation}
\mathbf{J}^{\rm sol} = \mathbf{I}\ .
\end{equation}
After the collective coordinate rotation, the total spin of the light degrees of freedom in the heavy baryon is defined
as
\begin{eqnarray}
\mathbf{j} & = & \mathbf{J}^{\rm sol} + \mathbf{g} 
= \mathbf{J}^{\rm sol} + \mathbf{r} + \mathbf{K}.
\label{def j}
\end{eqnarray}
By including the heavy quark spin, the spin operator for the heavy
baryon is expressed as $\mathbf{J}_B = \mathbf{j} \pm \mathbf{S}_Q$
with eigenvalues $ j_{B,\pm} = j \pm 1/2$ (in the case of $j = 0$,
only $j_B = 1/2$ exists). Then, we can express the collectively
rotated Hamiltonian as
\begin{eqnarray}
H_{\rm coll} & = & \frac{1}{2\mathcal{I}}\Big[[1-\chi(k)]\mathbf{I}^2 +
\chi(k)[\chi(k)-1]{\bf K}^{2}  \nonumber\\
& &  {} + 
\chi(k)(\mathbf{j} - \mathbf{r})^2
\Big]
\ .\label{eq:collechamilt}
\end{eqnarray}

Gathering all the above contributions, we finally obtain the heavy
baryon mass as
\begin{eqnarray}
M_{\rm(heavy\ baryon)} & = & M_{\rm sol} + \bar{M}_{H,G} + V_{H,G} + H_{\rm coll},
\label{mass formula}
\end{eqnarray}
where $V_{H}$ ($V_G$) is the binding energy corresponding to the
heavy meson $H$ ($G$). 
$\bar{M}_{H,G}$ are the weight-averaged heavy meson masses
with $\bar{M}_H = (3m_{D^{\ast}}+m_D)/4$ and $\bar{M}_G =
(3m_{D_1}+m_{D_0^{\ast}})/4$. Note that each combination of 
$(I,j)$ generates a pair of degenerate states with $ j_{B,\pm} = j
\pm 1/2$.

\section{Chiral Doubling of Heavy Baryons}

\label{sec:heavybaryon}

In this section, we study the chiral doubling structure
using the formulas obtained in the previous section.
In the following analysis, we restrict ourselves to the case with
$\mathbf{r} = 0$ and use the following values of heavy meson masses as inputs:
\begin{eqnarray}
\left( m_D,m_{D^\ast} , m_{D^\ast_0} , m_{D_1} \right)
=
\left( 1.86 , 2.01 , 2.32 , 2.42 \right)
~ \mbox{[GeV]} , \nonumber\\
\end{eqnarray}
which lead to
\begin{eqnarray}
\left(\bar{M}_H \,,\, \bar{M}_G \right) = 
\left(1.97 \,,\, 2.40 \right) \ \mbox{[GeV]}
\ .
\end{eqnarray}
Furthermore, to simulate the profile functions $F(r)$ and $\omega(r)$, which are necessary to evaluate the binding energy $V_H$ and $V_G$ expressed in Eq.~(\ref{eq:bindenergy}), we use the following inputs:
\begin{eqnarray}
\left( m_N\,,\,m_\Delta\right) = \left(0.94 \,,\,1.23\right)
\ \mbox{[GeV]}
\ , \label{eq:inputbaryonmass}
\end{eqnarray}
which yield soliton mass $M_{\rm sol} = 0.868$~GeV and the inverse of the moment of inertia $1/\mathcal{I} = 0.193$~GeV. Then, using the relevant expressions from HLS up to $\mathcal{O}(p^4)$ including the hWZ terms given in Refs.~\cite{MOYHLPR,MOYH}, by taking $F_\pi$ and $m_\rho$ as free parameters to fit the inputs (\ref{eq:inputbaryonmass}), we obtain $F_\pi = 62.24~$MeV and $m_\rho = 417.5~$MeV and the values of the profile functions at origin as 
\begin{eqnarray}
F^\prime(0) = 626.1{\rm ~MeV} ; \qquad \omega(0) = - 74.5{\rm ~MeV}. \label{eq:profile}
\end{eqnarray}
Moreover, we take the parameter $a = 2$~\cite{Harada:2003jx,Bando:1984ej} and 
fix the universal coupling constant $g$ in HLS through
\begin{eqnarray}
g = m_\rho/(F_\pi \sqrt{a}) = 4.74. \label{eq:numbg}
\end{eqnarray}

Let us first consider the binding energy in order to determine which channel can form the bound state and
calculate the spectra of the chiral partners. Using Eq.~(\ref{eq:bindenergy}), we obtain the binding energy between the heavy-light mesons in the $H$ doublet and
soliton as
\begin{eqnarray}
V_{\rm H} & = & {} - 0.177 \left(1+\kappa\right) + 0.626 g_A \left[k(k+1) - \frac{3}{2}\right] \mbox{[GeV]}.\nonumber\\ \label{eq:bindingenergyH}
\end{eqnarray}
The value of $g_A$ is determined through the $D^{\ast} \to D \pi$ decay as
$\vert g_A \vert = 0.56$~\cite{HHM1,HHM2}. It does not seem possible to determine 
the $\omega$ coupling constant $\kappa$ from the available
experimental data for the heavy meson decay.
In the case of the vector meson dominance we have $\kappa = 0$; therefore, it is reasonable
to regard $|\kappa| \lesssim 1$. Then we conclude that the $k=0$
channel gives a bound state when $g_A > 0$.
This bound state can be naturally identified with 
$\Lambda_c (\frac{1}{2}^+\,,\,2286)$
from the quantum number so that we assume $g_A > 0$ in the following analysis.
Since the collective energy is zero for the $k = 0, I = 0$ state,
we use the experimental value of the mass of
$\Lambda_c (\frac{1}{2}^+\,,\,2286)$ as an 
input to determine the value of $\kappa$.
Using $M_{\Lambda_c} = 2.29$\,GeV, we obtain $\kappa = - 0.83$.

Next, we consider the bound states made from the $G$ doublet. The binding energy is expressed as
\begin{eqnarray}
V_{\rm G} & = & {} - 0.177 \left(1+\kappa\right) 
 - 0.626 g_A \left[k(k+1) - \frac{3}{2}\right]~\mbox{[GeV]} . \nonumber\\
\end{eqnarray}
As we can see easily, $V_G > 0$ for $k = 0$ ($g_A > 0$) so that there is no bound state in the $k = 0$ channel. For the $k = 1$ channel, using $\kappa ={} - 0.83$ determined above, 
we obtain $V_G ={} - 0.205\,$GeV, which implies that the $k = 1$ channel is actually bound. 

From Eq.~(\ref{def j}) the total spin of the light degrees of
freedom becomes $1 ~(I =0)$ so that the resultant heavy baryons form a heavy-quark doublet 
consisting of $\Lambda_c(\frac{1}{2}^-)$ and $\Lambda_c(\frac{3}{2}^-)$. 
Combined with the collective energy, 
the mass of the bound state is expressed as
\begin{eqnarray}
M_{\Lambda_c(\frac{1}{2}^-,\frac{3}{2}^-)} = M_{\rm sol} + \bar{M}_G + V_G 
+ \frac{1}{4 {\mathcal I} }
\ , 
\end{eqnarray}
which leads to
\begin{eqnarray}
M_{\Lambda_c(\frac{1}{2}^-,\frac{3}{2}^-)} = 3.13\,\mbox{[GeV]} \ .
\end{eqnarray}
This value is much larger than the experimental values 
for the masses of negative parity baryons:
$M(\Lambda_c(\frac{1}{2}^-\,,\,2595) = 2.59\,$GeV and
$M(\Lambda_c(\frac{3}{2}^-\,,\,2625) = 2.63\,$GeV.
So, we conclude that the negative parity baryons found by experiments 
$\Lambda_c(\frac{1}{2}^-\,,\,2595)$ and $\Lambda_c(\frac{3}{2}^-\,,\,2625)$ are not 
the chiral partner to the ground state baryon $\Lambda_c(\frac{1}{2}^+\,,\,2285)$. In the present bound state approach, they should be regarded as the $r = 1$ state made from the $H$ doublet and nucleon.
Then, we expect to have a doublet for the chiral partner 
around the $3.1$\,GeV region.

We next study the $I=1$ baryons. In the positive parity baryon sector, the $k=0$ channel is bound so that
the total spin of the light degrees of freedom in the heavy baryon
becomes $1$.  As a result, the spin of $\Sigma_c$ baryons 
with positive parity is either $1/2$ or $3/2$.
In the mass formula in Eq.~(\ref{mass formula}), 
only $H_{\rm coll}$ changes its value depending on the isospin of baryons.
Since $\chi(k)=0$ for $k=0$, the mass difference between 
the $\Sigma_c$ and the $\Lambda_c$ in the positive negative parity
is obtained as
\begin{eqnarray}
M_{\Sigma_c(\frac{1}{2}^+,\frac{3}{2}^+) } - M_{\Lambda_c(\frac{1}{2}^+)} = \frac{1}{\mathcal I}
\ .
\end{eqnarray}
In the negative parity baryon sector, the $k=1$ channel is bound, then the
eigenvalue for ${\mathbf j}$ is either $0$, $1$, or $2$. We summarize our predicted results for the charm baryon spectrum in Tables~\ref{table:quantumnumbersch} and ~\ref{table:quantumnumberscg}.

We next study the mass spectrum of the bottom baryon by
substituting the bottom meson masses into the charm meson
masses in Eq.~(\ref{mass formula}).
In the bottom meson spectrum, the masses of the ground states $B$ and
$B^{\ast}$ are well measured but masses of the mesons in the $G$
doublets are not well established. Here, we naively estimate them
using $m_{B_0^{\ast}} - m_B = m_{D_0^{\ast}} - m_D = 2403 - 1869.6 =
533.4~$MeV and $m_{B_1^{\ast}} - m_{B^{\ast}} = m_{D_1^{\ast}} -
m_{D^{\ast}} = 2427 - 2010.25 = 416.75~$MeV which lead to
$m_{B_0^{\ast}} = 5812.9~$MeV and $m_{B_1^{\ast}} = 5741.85~$MeV.
Our numerical results for the masses of the heavy baryons including
the bottom quark with the corresponding quantum numbers are given in
Tables~\ref{table:quantumnumbersboth} and~\ref{table:quantumnumbersbotg}.
\begin{table}%
\caption{Predicted mass for the charmed baryon for the $H$ doublet. \label{table:quantumnumbersch}}
\begin{tabular}{lllllll}
\hline \hline  $I$ \,\,\,\,\, & $j$ \, & \,\,\,\,\, states \, & \,\,\,\,\, $M^H$(MeV) \\
\hline
 $0$ \,\,\,\,\, & $0$ & \,\,\,\,\, $\Lambda_c(\frac{1}{2}^+)$ & \,\,\,\,\, $2286.46$(input) \\
\hline
 $1$ \,\,\,\,\, & $1$ & \,\,\,\,\, $\Sigma_c(\frac{1}{2}^+),\Sigma_c(\frac{3}{2}^+)$ & \,\,\,\,\, $2481.13$ \\
\hline \hline
\end{tabular}
\end{table}
\begin{table}%
\caption{Predicted mass for the charmed baryon for the $G$ doublet. \label{table:quantumnumberscg}}
\begin{tabular}{llllll}
\hline \hline  $I$ \,\,\,\,\, & $j$ \, & \,\,\,\,\, $I(j_B^P)$ \, & \,\,\,\,\, $M^G$(MeV) \\
\hline
 $0$ \,\,\,\,\, & $1$ & \,\,\,\,\, $\Lambda_c(\frac{1}{2}^-),\Lambda_c(\frac{3}{2}^-)$ & \,\,\,\,\, $3131.66$ \\
\hline
 $1$ \,\,\,\,\, & $0$ & \,\,\,\,\, $\Sigma_c(\frac{1}{2}^-)$ & \,\,\,\,\, $3131.66$ \\
\hline
 $1$ \,\,\,\,\, & $1$ & \,\,\,\,\, $\Sigma_c(\frac{1}{2}^-),\Sigma_c(\frac{3}{2}^-)$ & \,\,\,\,\, $3228.99$ \\
\hline
 $1$ \,\,\,\,\, & $2$ & \,\,\,\,\, $\Sigma_c(\frac{3}{2}^-),\Sigma_c(\frac{5}{2}^-)$ & \,\,\,\,\, $3423.66$ \\
\hline \hline
\end{tabular}
\end{table}

\begin{table}%
\caption{Predicted mass for the bottom baryon for the $H$ doublet. \label{table:quantumnumbersboth}}
\begin{tabular}{lllll}
\hline \hline  $I$ \,\,\,\,\, & \, $j$ \, & \,\,\,\,\, $I(j_B^P)$ \,& \,\,\,\,\, $M^H$(MeV) \\
\hline
 $0$ \,\,\,\,\, & \, $0$ \, & \,\,\,\,\, $\Lambda_b(\frac{1}{2}^+)$ & \,\,\,\,\, $5625.07$ \\
\hline
 $1$ \,\,\,\,\, & \, $1$ \, & \,\,\,\,\, $\Sigma_b(\frac{1}{2}^+),\Sigma_b(\frac{3}{2}^+)$ \, & \,\,\,\,\, $5819.74$ \\
\hline \hline
\end{tabular} 
\end{table}
\begin{table}%
\caption{Predicted mass for the bottom baryon for the $G$ doublet. \label{table:quantumnumbersbotg}}
\begin{tabular}{lllll}
\hline \hline  $I$ \,\,\,\,\, & $j$ \, & \,\,\,\,\, $I(j_B^P)$ \, & \,\,\,\,\, $M^G$(MeV) \\
\hline
 $0$ \,\,\,\,\, &  $1$  & \,\,\,\,\, $\Lambda_b(\frac{1}{2}^-),\Lambda_b(\frac{3}{2}^-)$ \, & \,\,\,\,\, $6470.27$  \\
\hline
 $1$ \,\,\,\,\, & $0$ \, & \,\,\,\,\, $\Sigma_b(\frac{1}{2}^-)$ & \,\,\,\,\, $6470.27$  \\
\hline
 $1$ \,\,\,\,\, & $1$ \, & \,\,\,\,\, $\Sigma_b(\frac{1}{2}^-),\Sigma_b(\frac{3}{2}^-)$ & \,\,\,\,\, $6567.6$  \\
\hline
 $1$ \,\,\,\,\, & $2$ \, & \,\,\,\,\, $\Sigma_b(\frac{3}{2}^-),\Sigma_b(\frac{5}{2}^-)$ & \,\,\,\,\, $6762.27$  \\
\hline \hline
\end{tabular}
\end{table}

\section{Pentaquarks with heavy quark}

\label{sec:penta}

We next consider the pentaquark channel. Although the existence of these kinds of states still needs experimental confirmation, theoretical study of them is meaningful.
For a pentaquark state, the large component of the antiheavy quark
can be projected out with the projection operator
$(1-v\hspace{-0.2cm}\slash)/2$ so that in case the heavy meson is
at rest, the $H$ doublet has nonvanishing elements only in the
lower-left $2 \times 2$ sub-block while the $G$ doublet has
nonvanishing elements only in the lower-right $2 \times 2$ sub-block,
i.e.,
\begin{eqnarray}
H & = & \left(
          \begin{array}{cc}
            0 & 0 \\
            \mathbf{H} & 0 \\
          \end{array}
        \right)
, G = \left(
          \begin{array}{cc}
            0 & 0 \\
            0 & \mathbf{G} \\
          \end{array}
        \right) , \nonumber\\
\bar{H} & = & \gamma_0 H^\dag \gamma_0 = \left(
          \begin{array}{cc}
            0 & -\mathbf{H}^\dag \\
            0 & 0 \\
          \end{array}
        \right), \nonumber\\
\bar{G} & = & \gamma_0 G^\dag \gamma_0 = \left(
          \begin{array}{cc}
            0 & 0 \\
            0 & \mathbf{G}^\dag \\
          \end{array}
        \right).
\end{eqnarray}

Then, substituting $v_\mu$ with $-v_\mu$, following the above
derivation, one can see that both the binding energies given by
Eq.~(\ref{eq:bindenergy}) change a sign. As a result, the binding energies for the pentaquark states made from the anti-$H$ doublet and the pentaquark states made from the anti-$G$ doublet are expressed as
\begin{eqnarray}
V_{\rm H}^5 & = & {} 0.177 \left(1+\kappa\right) - 0.626 \, g_A \left[k(k+1) - \frac{3}{2}\right] \quad \mbox{[GeV]} , \nonumber\\
V_{\rm G}^5 & = & {} 0.177 \left(1+\kappa\right) + 0.626  g_A \left[k(k+1) - \frac{3}{2}\right]\quad \mbox{[GeV]}
\ . \label{eq:bindingEH5}
\end{eqnarray}
Therefore, for the anti-$H$ doublet, the $k=1$ channel gives the bound states with binding energy $V_{\rm H}^5 = - 145.6$~MeV, while for anti-$G$ doublet, the $k=0$ channel gives the bound states with binding energy $V_{\rm G}^5 = - 496.2$~MeV.

Substituting relevant numerical results, we obtain the spectrum of the
pentaquark states. We list our results in
Tables~\ref{table:quantumnumberspentach} and~\ref{table:quantumnumberspentacg} for pentaquark states consisting
anti-charm quark and Tables~\ref{table:quantumnumberspentabh} and~\ref{table:quantumnumberspentabg} for
pentaquark states consisting anti-bottom quark.
\begin{table}%
\caption{Predicted mass for the charmed pentaquark state for the $H$
doublet.
\label{table:quantumnumberspentach}}
\begin{tabular}{llllll}
\hline \hline  $I$ \, & $j$ \, & Candidates \, & $M^{5,H_c}$(MeV) \\
\hline
 $0$ & $1$ \, & $\Theta_c(\frac{1}{2}^+)$ \, & $2745.15$ \\
\hline \hline
\end{tabular}
\end{table}
\begin{table}%
\caption{Predicted mass for the charmed pentaquark state for the $G$ doublet.
\label{table:quantumnumberspentacg}}
\begin{tabular}{llllll}
\hline \hline  $I$ \, & $j$ \, &  Candidates \, & $M^{5,G_c}$(MeV) \\
\hline
 $0$ & $0$ \, & $\Theta_c(\frac{1}{2}^-)$ \, & $2791.78$ \\
\hline \hline
\end{tabular}
\end{table}
\begin{table}%
\caption{Predicted mass for the bottom pentaquark state for the $H$
doublet.
\label{table:quantumnumberspentabh}}
\begin{tabular}{lllll}
\hline \hline  $I$ \, & $j$  \, & Candidates \, & $M^{5,H_b}$(MeV) \\
\hline
 $0$ & $1$ \,& $\Theta_b(\frac{1}{2}^+)$  \, & $6083.76$ \\
\hline \hline
\end{tabular} 
\end{table}
\begin{table}%
\caption{Predicted mass for the bottom pentaquark state for the $G$ doublet.
\label{table:quantumnumberspentabg}}
\begin{tabular}{lllll}
\hline \hline  $I$ \, & $j$ \, & Candidates \, & $M^{5,G_b}$(MeV) \\
\hline
 $0$ & $0$ \, & $\Theta_b(\frac{1}{2}^-)$ \, & $6130.39$ \\
\hline \hline
\end{tabular}
\end{table}

The results in Tables~\ref{table:quantumnumberspentach} and~\ref{table:quantumnumberspentacg} show that the
lightest charmed pentaquark states made of soliton and heavy-light
mesons in the anti-$H$ doublet have masses of about $2.75~$GeV,
and their chiral partner made of the anti-$G$ doublet
has a mass of about $2.80~$GeV. Both of them are below the $Dp$
threshold.  The reason that the pentaquark states from the anti-$H$
doublet and anti-$G$ doublet have similar masses is because the
binding energy of the anti-$H$ doublet is about $350$~MeV smaller than
that of the anti-$G$ doublet, and the collective rotation energy, which is
about $50$~MeV, does not contribute to the latter. With respect to the
status of the pentaquark search performed, these states cannot be 
ruled out, and since they cannot decay via a strong process, their total
widths should be narrow.

\section{Summary and Discussions}

\label{sec:conclusion}

We studied the chiral partner structure of heavy baryons in the
bound state approach including the vector meson exchanging effects
through the hidden local symmetry. We showed that in the large $N_c$
limit and the heavy quark limit, the ground state heavy baryon made of the
ground state heavy-light meson and the nucleon has a chiral
partner made of an excited heavy-light meson and nucleon. Our
explicit calculation showed that the chiral partner of
$\Lambda_c(\frac{1}{2}^+)$ is a heavy quark doublet of $\Lambda(\frac{1}{2}^-)$ and $\Lambda(\frac{3}{2}^-)$.
This contrasted to the perdition made in the pioneering work in 
Ref.~\cite{Oh:1994ky}, where the chiral partner was the singlet under the heavy quark spin
transformation. Our prediction of the mass was about $3.1$\,GeV, 
which indicated that the $\Lambda_c(\frac{1}{2}^-,2595)$ and $\Lambda_c(\frac{3}{2}^-,2625)$
listed in the PDG table~\cite{Nakamura:2010zzi} should be
interpreted as the $r=1$ excitation of $\Lambda_c(\frac{1}{2}^+)$. To calculate the spectrum of the $r \ne 0$ states, one should consider the relative motion of the soliton with respect to the heavy mesons~\cite{Oh:1997tp}. This is beyond the scope of the present paper.

We also studied the bound states in the pentaquark channel. We found that the $k = 1$ channel forms bound states for 
the anti-$H$ doublet ($\Theta_c(\frac{1}{2}^-)$, $\Theta_c(\frac{3}{2}^-)$), 
while the $k = 0$ channel forms bound states for the anti-$G$ doublet
($\Theta_c(\frac{1}{2}^+)$). It was found that the predicted masses of the pentaquark states made of the
anti-$H$ doublet and anti-$G$ doublet were below the $Dp$ threshold, which cannot be ruled out by the present data~\cite{Pentadata}.

In the present analysis, we took the infinite
heavy soliton and heavy quark limits so that both the soliton and
heavy-light meson were sitting at the origin.
This picture could not be applied to the bound states with nonzero $r$.
Since in the present analysis, the chiral
partner of heavy $\Lambda_c(\frac{1}{2}^+,2286)$ had a mass of
about $3.1$\,GeV, which was a bound state of soliton and heavy-light
mesons in the $G$ doublet, one could expect that it had broad width due
to the broad width of the constituent $P$-wave mesons in the $G$ doublet.
From the numerical results in Tables~\ref{table:quantumnumbersboth} and~\ref{table:quantumnumbersbotg} we
concluded that the spectrum of the heavy baryons with bottom quark was
consistent with PDG~\cite{Nakamura:2010zzi} for $\Lambda_b$ and
$\Sigma_b$.

It should be noted that in the present analysis, we considered that
the chiral partner to the nucleon was itself: The left-handed nucleon
was the chiral partner of the right-handed nucleon, and vice versa so that the chiral partner to the heavy baryon as the bound state of
the $H$ doublet and the nucleon was the one made of the $G$ doublet and
the nucleon. This implied that the chiral partner structure of the
heavy baryons in our approach arose from the chiral partner
structure of the constituent heavy-light mesons. On the other hand,
in the mirror scenario for the light baryon~\cite{mirror}, the
chiral partner to the nucleon was considered as $N(1535)$. In such a
case, the full picture of the chiral partner structure of heavy
baryons became complicated, and we did not consider this scenario in
the present work.

\appendix

\section{ Matrix element of heavy-light meson isospin operator}

\label{app:kl}

Using the Wigner-Eckart theorem, we can express the matrix element
of heavy-light meson isospin operator $I^a_{\rm light}$ in terms of
the matrix element of the operator $\mathbf{K}$, i.e.,
\begin{widetext}
\begin{eqnarray}
\int d\Omega\langle \psi_{gg_3}^{(i)}|\mathbf{I}_{\rm light}|
\psi_{gg_3}^{(j)}\rangle & = & \int d\Omega \langle
\psi_{gg_3}^{(i)}|\mathbf{K}| \psi_{gg_3}^{(j)}\rangle \frac{\langle
\psi_{gg_3}^{(i)}|\mathbf{K} \cdot \mathbf{I}_{\rm light}|
\psi_{gg_3}^{(j)}\rangle }{k(k+1)} \nonumber\\
& = & \int d\Omega \frac{\langle \psi_{gg_3}^{(i)}|[\mathbf{K}^{2} +
\mathbf{I}^2 - \mathbf{J}_{\rm light}^2 ]| \psi_{gg_3}^{(j)}\rangle
}{2k(k+1)} \langle \psi_{gg_3}^{(i)}|\mathbf{K}|
\psi_{gg_3}^{(j)}\rangle
\nonumber\\
& = & \frac{[k(k + 1) + 3/4 - j_l(j_l+1)]}{2k(k+1)} \int d\Omega
\langle \psi_{gg_3}^{(i)}|\mathbf{K}| \psi_{gg_3}^{(j)}\rangle
\delta_{ij} \nonumber\\
& = & \chi(k) \mathbf{K},
\end{eqnarray}
\end{widetext}

\acknowledgments

\label{ACK}

We are grateful to Y.~Oh, B.-Y. Park, and M.~Rho for valuable comments and critical reading of 
the manuscript. This work is supported in part by Grant-in-Aid for Scientific
Research on Innovative Areas  Grant No. 2104, ``Quest on New Hadrons with
Variety of Flavors'' from MEXT. The work of M.H. is supported in
part by the Grant-in-Aid for Nagoya University Global COE Program
``Quest for Fundamental Principles in the Universe: From Particles
to the Solar System and the Cosmos'' from MEXT, the JSPS
Grant-in-Aid for Scientic Research Grabts No.(C) 24540266 and No. (S) 22224003.
The work of Y.M. is supported in part by the
National Science Foundation of China (NSFC) under Grant No.
10905060.


\end{document}